\documentclass{aa}
\usepackage{epsf}
\begin{document}
\sloppypar

%
   \title{Reflection and noise in the low spectral state of GX~339-4}

   \author{M. Revnivtsev\inst{1,2}, M. Gilfanov\inst{2,1},
   E. Churazov\inst{2,1}
	}

   \offprints{revnivtsev@hea.iki.rssi.ru}

   \institute{Space Research Institute, Russian Academy of Sciences,
              Profsoyuznaya 84/32, 117810 Moscow, Russia,
        \and
                Max-Planck-Institute f\"ur Astrophysik,
              Karl-Schwarzschild-Str. 1, 85740 Garching bei M\"unchen,
              Germany
             }
  \date{}

	\authorrunning{Revnivtsev, Gilfanov \& Churazov}
	\titlerunning{Reflection and noise in the low spectral state of
	GX~339-4}
	
   \abstract{
	We analyze RXTE/PCA observations of GX339-4 in the low spectral state
	from 1996--1997 and show that the pattern of its spectral and temporal
	variability is nearly identical to that of Cyg X-1. In particular, a
	tight correlation exists between the QPO centroid frequency and the
	spectral parameters. An increase of the QPO centroid frequency is
	accompanied with an increase of the amplitude  of the reflected component
	and a steepening the slope of the underlying power law.
	Fourier frequency resolved spectral analysis showed, that the  
	variability of the reflected component at frequencies
	higher than $\sim 1-10$ Hz is suppressed in comparison with that
	of the primary emission. 
      \keywords{accretion, accretion disks -- black hole physics -- stars:
               binaries: general -- stars: individual:(GX339-4) --
               X-rays: general -- X-rays:stars
               }
}

\maketitle
%

\section{Introduction}

It was found recently  for a large sample of Seyfert AGNs and several
observations of the Galactic X-ray binaries that the amplitude of the
reflected component is generally correlated with the slope of the
primary power law emission (\cite{zdz1}). Based on the numerous 
RXTE/PCA observations of Cyg X-1 \cite{pap2} (hereafter Paper I)
showed that this correlation is strong for multiple observations of 
this source and that the spectral parameters are also tightly
correlated with the characteristic noise frequency. In particular an
increase of the QPO centroid frequency is  accompanied with a
steepening of the slope of the Comptonized radiation and an increase
of the amplitude of the reflected component. Studying fast variability
of the reflected emission \cite{pap1} showed that its amplitude is
suppressed with respect to that of the primary emission at the
frequencies higher than $\sim 1-10$ Hz.

GX339-4 is a bright and well studied  X-ray binary. It is
usually classified as a black hole candidate and in many aspects is
very similar to Cyg X-1 (see e.g. \cite{tanakalewin}, \cite{tsp_339},
\cite{zdz}, \cite{wilms_339}, \cite{nowak_339}). The investigations 
of the connections between the spectral and timing properties of Cyg X-1
(e.g. \cite{pap2}) and GX 339-4 (e.g. \cite{ueda94}) indicate that 
these sources could be similar from this point of view also.
In this paper we expand the analysis of correlations between spectral and
temporal characteristics of the X-ray emission of GX 339-4 in the low
spectral state using the Rossi X-ray Timing Explorer data 
and show that this source demonstrates the same behavior that was
 previously observed from Cyg X-1.

\section{Observations, data reduction and analysis}

We used the publicly available data of GX 339-4 observations with RXTE/PCA
from 1996--1997 performed during the low spectral state of the
source. Our sample includes 23 observations from the proposals 10068,
20056, 20181 and 20183 with a total exposure time of $\sim$130 ksec
(Table \ref{gx339_pars_table}). Only observations from the proposal
P20183 had sufficient energy and timing resolution to perform
Fourier frequency resolved spectral analysis. Therefore the frequency
resolved analysis was carried out only for $\sim61$ ksec of the data.

The data screening was performed following the RXTE GOF
recommendations: offset angle $<0.02^\circ$, Earth 
elevation angle $> 10^\circ$, electron contamination value (the
``electron ratio'') for any of PCUs $<0.1$. 
The data from all PCUs were used for the analysis. 
The energy spectra were extracted from the PCA mode ``Standard 2''
(128 channels, 16 sec time resolution) and averaged over each
observation.
Fourier frequency resolved spectral analysis used ``Good Xenon'' data
(256 energy channels, 1$\mu$s time resolution). 
The response matrixes were built using standard RXTE FTOOLS 4.2
tasks (Jahoda 1999). The background spectra for the conventional
spectral analysis were constructed with the help of the ``VLE'' 
based model (Stark 1999). The background contribution to the frequency
resolved spectra   is negligible in the frequency and energy ranges of
interest.  A uniform systematic uncertainty of 0.5\% was added
quadratically to the statistical error in each energy channel. The
value of systematic uncertainty was chosen basing on the deviations
of the PCA Crab spectra from a power law model (see
e.g. \cite{wilms_339}).

\begin{figure}[t]
\epsfxsize=8cm
\epsffile[10 160 580 700]{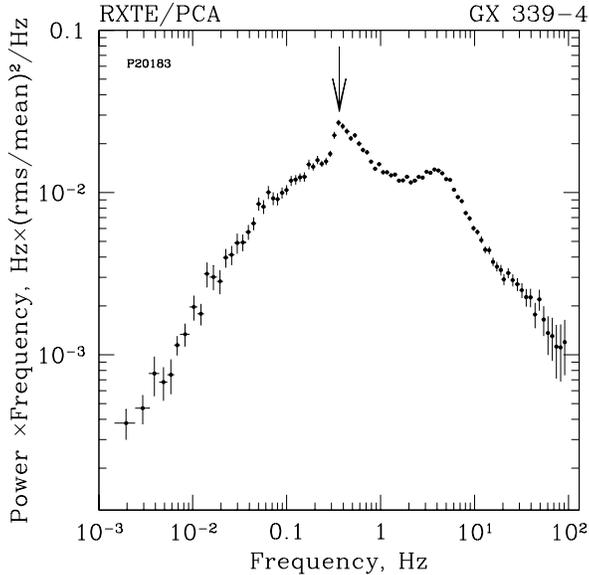}	
\caption{Typical power density spectrum  of GX 339-4 in
the low spectral state. The QPO peak at the frequency of $\sim$0.35 Hz
is shown by the arrow. In order to improve the statistics 
the power density spectrum was averaged over all observations of
the proposal P20183 having somewhat different QPO centroid frequency,
resulting in a somewhat broader profile of the QPO peak.  
 \label{gx339_power}}	
\end{figure}

\begin{figure*}[t]

\hbox{
\epsfxsize 8.5cm
\epsffile[10 150 580 690]{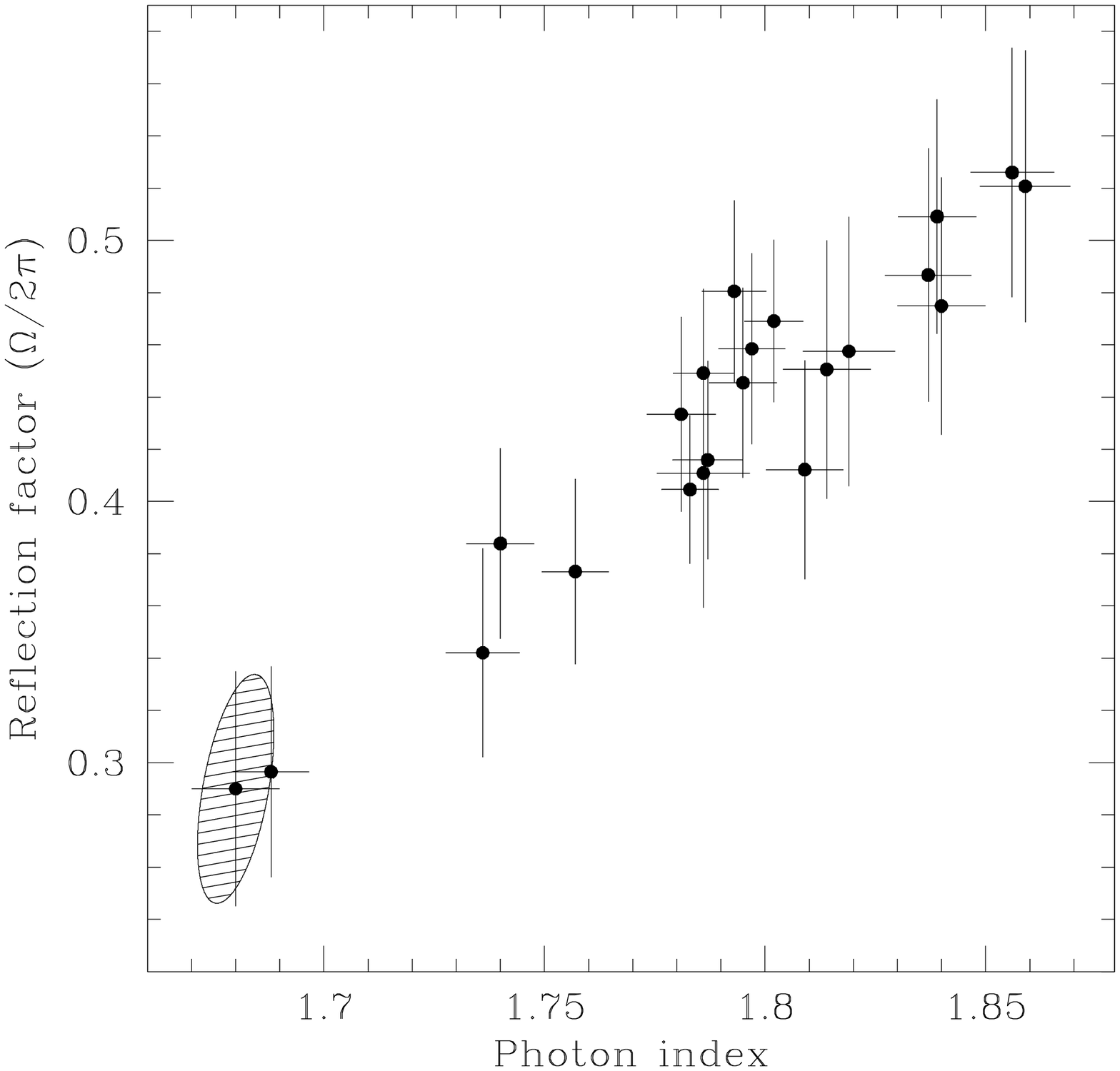}	
\epsfxsize 8.5cm
\epsffile[10 150 580 690]{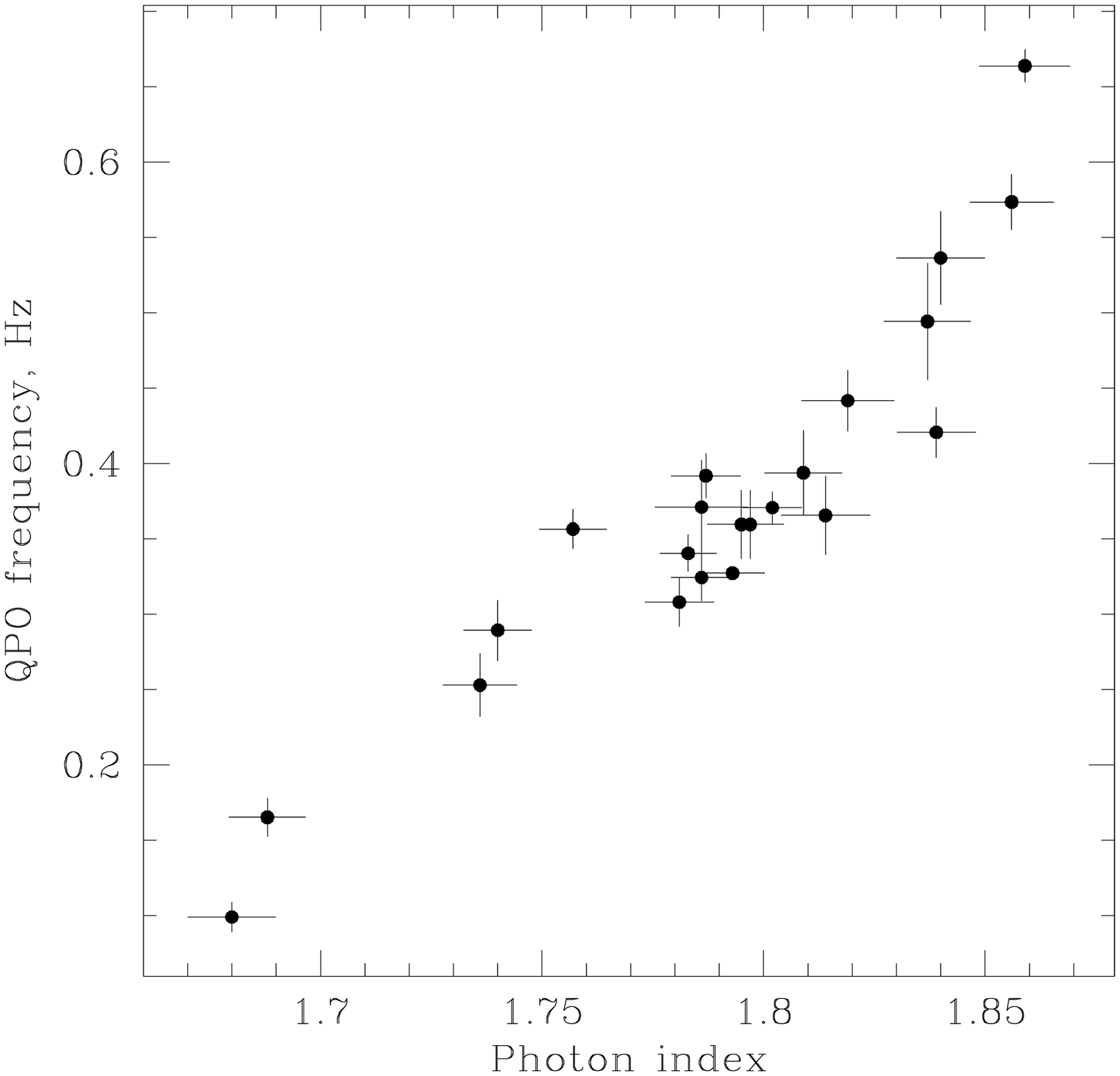}
}

\caption{The dependences of reflection scaling factor and the QPO
frequency on the slope of the primary power law. The filled oval shaped 
region on the left panel shows a typical 1-$\sigma$
confidence contour.\label{corr}} 
\end{figure*}

The energy spectra were fit in the 3--20 keV energy range with a
spectral model identical to that of Paper I. The model consisted of a
power law without high energy cutoff with superposed  continuum,
reflected from the  neutral medium ($pexrav$ model in XSPEC, see
\cite{m_zdz}) and an intrinsically  narrow emission line at the energy
6.4 keV. No ionization effects were taken into account.
The binary  system inclination angle was fixed at $\theta=45^o$
(e.g. \cite{zdz}), the iron abundance -- at the solar value.
In such a model the amplitude of the reflected component is
characterized by the reflection scaling factor $R$, which is an
approximate measure of the solid angle subtended by the reflector,
$R\sim\Omega/2\pi$. In the simplest geometry of an isotropic point
source above the infinite reflecting plane, the reflection scaling
factor $R$ is equal to 1. In order to approximately account for
smearing of the reflection features due to e.g relativistic and
ionization effects the reflected continuum  and the fluorescent line
were convolved with a Gaussian. Its width was a free parameter of the
fit. The uncertainties in Table \ref{gx339_pars_table} represents
1$\sigma$ confidence intervals for the model parameters. The error bars on
the values of equivalent width of the line were
calculated by the propagation of errors from line flux value.
   
The power spectra of GX339--4 in the low spectral state feature a
prominent QPO peak which frequency varies typically between $\sim 0.1$
and $\sim 0.5$ Hz (Fig.\ref{gx339_power}). We therefore used its
centroid to parameterize the characteristic noise frequency. The power
spectra were approximated with a model consisting of two band limited
noise components (Lorentzians, centered at zero frequency) and the
comparably narrow Lorentzian profile (QPO peak).

The Fourier frequency resolved spectra were obtained following
the prescription of Revnivtsev, Gilfanov \& Churazov, 1999 and were
approximated with the same model as averaged spectra except that the 
width of the Gaussian used to model the smearing of the reflection
features was fixed at the value of 0.7 keV.

\section{Results and their uncertainties}

\begin{table*}[t]
\caption{The list of observations and parameters of the energy and  power
density spectra approximation.\label{gx339_pars_table}}
\tabcolsep=0.17cm
\begin{tabular}{lccrcccccc} 
\hline 
Obs.ID&Date&Time,
UT&Exp.$^a$&$\Gamma$&$R\sim\Omega/2\pi$&EW, eV&$\sigma^b$, keV&$f_{QPO}, Hz$&$\chi^2(40_{\rm 
  dof})$\\
\hline
10068-05-01-00 & 17/10/96&02:12--05:07&  5550 &$1.68\pm0.01$&$0.29\pm0.04$&$143\pm30$&$0.64\pm0.09$&$0.09\pm0.01$&26.5 \\ 
10068-05-02-00 & 29/10/96&22:33--00:45&  5530 &$1.69\pm0.01$&$0.30\pm0.04$&$120\pm30$&$0.62\pm0.14$&$0.11\pm0.01$&37.6 \\ 
20056-01-01-00 & 05/04/97&08:36--09:15&  2118 &$1.79\pm0.01$&$0.41\pm0.05$&$173\pm40$&$0.85\pm0.11$&$0.37\pm0.03$&29.0 \\ 
20056-01-02-00 & 10/04/97&11:47--12:28&  2071 &$1.81\pm0.01$&$0.41\pm0.04$&$184\pm32$&$0.75\pm0.08$&$0.39\pm0.03$&38.7 \\ 
20056-01-03-00 & 11/04/97&13:25--14:06&  2122 &$1.81\pm0.01$&$0.45\pm0.05$&$202\pm38$&$0.83\pm0.09$&$0.36\pm0.03$&36.2 \\ 
20056-01-04-00 & 13/04/97&20:09--20:50&  2282 &$1.82\pm0.01$&$0.46\pm0.05$&$213\pm41$&$0.92\pm0.09$&$0.44\pm0.02$&32.5 \\ 
20056-01-05-00 & 15/04/97&20:41--21:20&  2049 &$1.84\pm0.01$&$0.47\pm0.05$&$218\pm37$&$0.89\pm0.08$&$0.53\pm0.03$&27.9 \\ 
20056-01-06-00 & 17/04/97&23:25--00:01&  2033 &$1.84\pm0.01$&$0.49\pm0.05$&$225\pm37$&$0.92\pm0.07$&$0.49\pm0.04$&34.2 \\ 
20056-01-07-00 & 19/04/97&22:20--23:08&  2045 &$1.86\pm0.01$&$0.53\pm0.05$&$216\pm33$&$0.79\pm0.07$&$0.57\pm0.02$&34.2 \\ 
20056-01-08-00 & 22/04/97&21:53--22:29&  1988 &$1.86\pm0.01$&$0.52\pm0.05$&$227\pm38$&$0.89\pm0.07$&$0.66\pm0.01$&41.7 \\ 
20181-01-01-01 & 03/02/97&15:56--19:09&  6622 &$1.80\pm0.01$&$0.46\pm0.04$&$191\pm30$&$0.84\pm0.07$&$0.34\pm0.02$&37.7 \\ 
20181-01-01-00 & 03/02/97&22:26--01:17&  5452 &$1.79\pm0.01$&$0.45\pm0.04$&$196\pm30$&$0.83\pm0.07$&$0.35\pm0.02$&40.6 \\ 
20181-01-02-00 & 10/02/97&15:49--20:22& 10535 &$1.79\pm0.01$&$0.48\pm0.03$&$190\pm29$&$0.86\pm0.07$&$0.32\pm0.01$&22.8 \\ 
20181-01-03-00 & 17/02/97&18:28--23:46& 11412 &$1.79\pm0.01$&$0.45\pm0.03$&$176\pm26$&$0.79\pm0.07$&$0.32\pm0.01$&24.3 \\ 
20183-01-01-01 & 08/02/97&14:20--20:25& 13702 &$1.80\pm0.01$&$0.47\pm0.03$&$185\pm26$&$0.80\pm0.06$&$0.37\pm0.02$&34.3 \\ 
20183-01-02-00 & 14/02/97&00:18--06:33&  9780 &$1.78\pm0.01$&$0.40\pm0.03$&$177\pm24$&$0.75\pm0.06$&$0.34\pm0.01$&21.4 \\ 
20183-01-02-01 & 14/02/97&14:20--21:22&  5779 &$1.78\pm0.01$&$0.43\pm0.04$&$182\pm30$&$0.81\pm0.08$&$0.30\pm0.02$&30.0 \\ 
20183-01-03-00 & 22/10/97&03:00--05:52&  6556 &$1.84\pm0.01$&$0.51\pm0.04$&$203\pm35$&$0.90\pm0.08$&$0.42\pm0.02$&38.5 \\ 
20183-01-04-00 & 25/10/97&03:22--06:00&  5385 &$1.79\pm0.01$&$0.42\pm0.04$&$160\pm28$&$0.72\pm0.09$&$0.39\pm0.01$&35.8 \\ 
20183-01-05-00 & 28/10/97&18:08--22:13&  6534 &$1.76\pm0.01$&$0.37\pm0.04$&$151\pm27$&$0.72\pm0.09$&$0.36\pm0.01$&28.2 \\ 
20183-01-06-00 & 31/10/97&19:41--22:10&  4621 &$1.74\pm0.01$&$0.34\pm0.04$&$129\pm29$&$0.63\pm0.12$&$0.25\pm0.02$&44.2 \\ 
20183-01-07-00 & 03/11/97&20:35--23:48&  7186 &$1.74\pm0.01$&$0.38\pm0.04$&$135\pm26$&$0.67\pm0.10$&$0.29\pm0.02$&30.7 \\ 
\hline
\end{tabular}
\begin{list}{}{}
\item[$^a$]-- Dead time corrected value
\item[$^b$]-- The width of the Gaussian used to model the smearing of
the reflection features.
\end{list}
\end{table*}

The results of the energy and power spectra approximation are
presented in Table \ref{gx339_pars_table} and Fig.\ref{corr}. 
As can be seen from Fig.\ref{corr} the main temporal and spectral
parameters -- characteristic noise frequency,
slope of the underlying power power law and amplitude of the reflected
component -- change in a correlated way. A steepening of the
spectrum is accompanied with an  increase of the reflection
amplitude and an increase of the QPO centroid frequency. Such a
behavior is very similar to that found in Paper I for Cyg X-1.

The spectral model is obviously oversimplified and does not include
several important effects such as ionization of the reflecting media,
deviations of the primary emission spectrum from the power law, exact 
shape of the relativistic smearing of the reflection features
etc. These effects might affect the
best fit parameters and could lead to
appearance of artificial correlations between them.
Particulary sensitive to the choice of the spectral model is the reflection
scaling factor $R\sim\Omega/2\pi$. As is well known there is some degeneracy
between the amplitude of reflection $R$ and the photon index $\Gamma$ of the
underlying power law determined from the spectral fits, especially if the
spectral 
fitting was done in a limited energy range (e.g. \cite{zdz1}). This
degeneracy might result in a 
slight positive correlation between the best fit values of $R$ and
$\Gamma$ which is in part due to statistical noise and, in
part, due to inadequate choice of the spectral model. The statistical
part of this degeneracy is illustrated in Fig.\ref{corr} by a
2-dimensional confidence contour for one of the points in the
$R-\Gamma$ plane. As can be easily seen from 
Fig.\ref{corr} it is correctly represented by the error bars assigned
to the points.

\begin{figure}
\epsfxsize=8cm
\epsffile{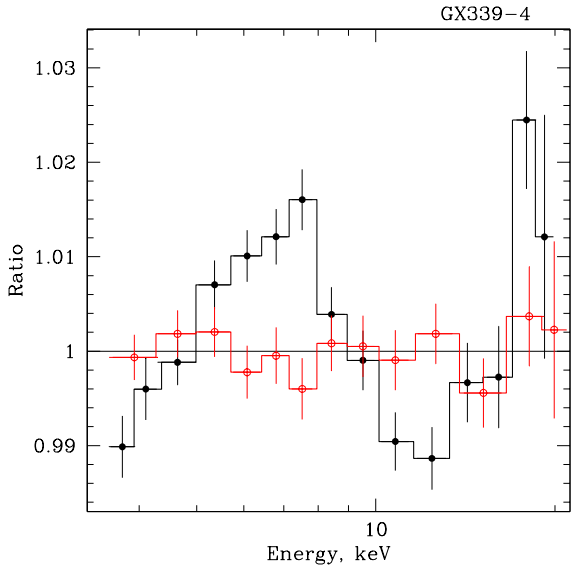}
\caption{Ratios of the observed counts spectra with different best fit
values of the reflection scaling factor $R$ and photon index
$\Gamma$. Solid circles show the ratio of the spectrum averaged over
 April 13--22, 1997 observations, having large best fit value of the
reflection ($R=0.5$, $\Gamma=1.84$) to the average spectrum of Oct. 17
and 19, 1996 with the smallest best fit value of reflection and
hardest power law photon index ($R=0.3$, $\Gamma=1.69$). The ratio was
multiplied by $E^{-0.09}$ for clarity. For comparison the open circles
show the ratio of the counts spectra of Oct. 25, 1997 and Feb. 14,
1997 having close values of the best fit parameters ($R=0.43$,
$\Gamma=1.78$ and $R=0.42$, $\Gamma=1.79$ respectively) and separated by the
similar period in time as the first ratio.
\label{ratios}} 
\end{figure}

In order to estimate contribution of the second, systematic
part of the $R-\Gamma$ degeneracy we compare two pairs of observations with
different and close best fit values of reflection factors $R$.
 We plot in Fig.\ref{ratios} the ratios of the count spectra for each
pair. As is clearly seen from Fig.\ref{ratios} the spectrum having
larger best fit value of the reflection scaling factor (and the
equivalent width of the line) shows more
pronounced reflection signatures -- the fluorescence line at $\sim
6.4$ keV followed by the 
absorption edge and increase due to the Compton reflected continuum at
larger energies. Thus we conclude  that although the best fit values of the
model parameters might not represent the exact values of  
the physically important quantities, our spectral model does
correctly rank the spectra according to the amplitude of the
reflection signatures and the correlations shown in Fig.\ref{corr}
are not artificial.

\begin{figure}[htb]
\epsfxsize 8 cm
\epsffile[10 170 580 670]{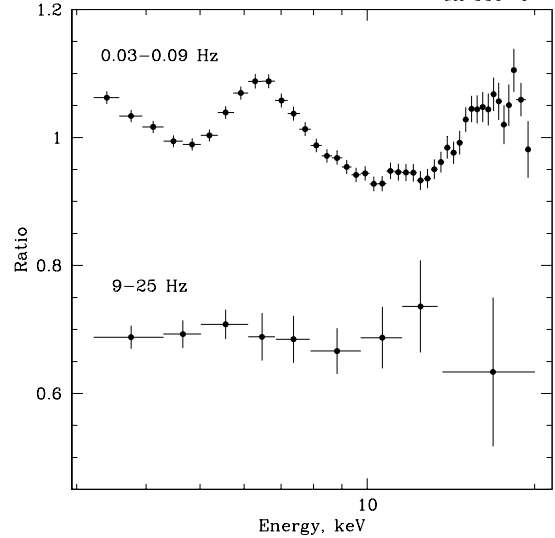}
\caption{The ratio of the frequency resolved spectra to a power law
model with photon index of $\alpha$=1.8. The 9-25 Hz spectrum 
was multiplied by 0.7 for clarity. 
\label{spectra}
}
\end{figure}

\begin{figure}[htb]
\epsfxsize 8.5 cm
\epsffile[10 170 580 710]{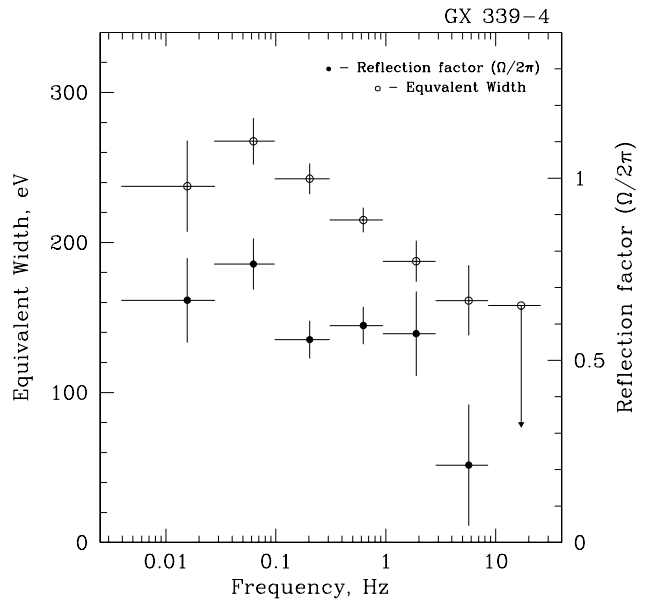}
\caption{The Fourier frequency dependence of the equivalent width of
the fluorescent line and the reflection scaling factor R.
The upper limit on the reflection scaling factor in the 9--25 Hz
frequency range equals to 1.5 and not shown in the graph.
\label{eqw}
}
\end{figure}

The Fourier frequency resolved spectra illustrate the energy
dependence of the amplitude of X-ray flux variation at a given time scale.
As was stressed out by \cite{pap1} and \cite{freqres2} interpretation
of the Fourier frequency resolved spectra in general is not
straightforward and requires certain {\em a priory} assumptions to be
made. One of the areas where it can be 
efficiently used and which at present can not be accessed by
the conventional spectroscopy is studying the fast variability of the
reflected emission. 
Indeed, variation of the parameters in the Comptonization region, for 
instance, lead to variations of the spectral shape of the
Comptonized radiation which would be imprinted in the 
Fourier frequency resolved spectra. Their shape, however, might 
differ significantly from any of the Comptonized spectra they resulted
from and no easily interpretable results could be obtained via
the conventional spectral fits. 
The shape of the reflection signatures, on the other hand, and
especially that of the fluorescent line, is generally  subject to
significantly less variations. Therefore they can be easily
identified in the frequency resolved spectra and their amplitude
can be measured. Absence or presence of the reflection signatures in a 
Fourier frequency resolved spectrum would signal absence or presence
of variations of the reflected emission at the given
frequency. Their amplitude would in principal measure
amplitude of the variations of the reflected flux relative to the
variations of the primary emission.   With that in mind we show in
Fig.\ref{spectra} the Fourier frequency resolved spectra of GX339--4
in two frequency ranges.  Significant decrease of the
amplitude of the reflection features with frequency is apparent. 
Quantitative dependence of the reflection amplitude on the Fourier
frequency is shown in Fig.\ref{eqw}. Within the available statistical
accuracy this dependence is  qualitatively and quantitatively similar
to that found for Cyg X-1 (\cite{pap1}).

\section{Discussion}

We analyzed 23 observations of GX339-4 with RXTE/PCA performed from
1996--1997 during the low spectral state of the source. Using simple
spectral model we found that
the pattern of temporal and spectral variability of GX339-4 is confirming
the previous findings of \cite{ueda94}, obtained with the help of GINGA
observatory, and it is nearly
identical to that of Cyg X-1 (\cite{pap1}, Paper I). This indicates that
such a pattern might be common for the accreting black holes in the
low spectral state.  In particular:

\begin{enumerate}

\item

The characteristic noise frequency, slope of the Comptonized spectrum
and amplitude of the reflected component change in a correlated
way. Increase of the noise frequency is accompanied by increase of the
amplitude of the reflected component and steepening of the
Comptonized spectrum. 

\item

Fourier frequency resolved spectral analysis showed that 
the short term variations of the reflected flux are suppressed in
comparison with variations of the primary Comptonized flux at
frequencies above $\sim 1-10$ Hz.
\end{enumerate}

\sloppypar 
As it was discussed e.g. in Gilfanov et al. 1999, \cite{zdz1}
 the correlation of the spectral
index (the photon index $\Gamma$ of the underlying power law) with the amplitude of the reflected component hints on the possible
relation between the solid angle, subtended by the reflector, and the 
influx of the soft photons to the Comptonization region. 
Existence of such relation suggests that the reflecting medium is 
the primary source of the soft seed photons to the Comptonizing region. 
It could be explained 
e.g. within the framework of the disk-spheroid models of the accretion flow 
(see e.g. \cite{poutanen97}) (a hot quasi-spherical  Comptonization
region near the compact object surrounded by an optically thick cold
accretion disk terminating at some radius $R_{\rm in}$). In such geometry the decrease of the inner radius of the 
optically thick cold accretion disk should result in: 1)
the decrease of the temperature of the inner hot region, because of
 the increase of the influx of cold seed photons, and 2) the increase of the 
solid angle subtended by this accretion disk (reflector) seen from the 
central hard source. This would lead to the correlation of the spectral
 index and the amplitude of the reflection. In addition, if we 
assume that the characteristic frequencies of aperiodic variations
of the source flux are proportional to the keplerian frequency on the inner 
boundary of the optically thick accretion disk we will obtain an additional 
correlation. The characteristic frequency of the source's power spectrum will
positively correlate with the amplitude of the reflection. This is  
what we observe in the cases of Cyg X-1 and GX 339-4. We should note,
 however, that the above explanation of the observed $R-\Gamma$ correlation 
is not unique. For example, the similar dependencies can be produced in the 
model of active corona above the accretion disk 
(see e.g. \cite{beloborodov99}, \cite{tim_review}). 

The frequency resolved spectral analysis was introduced in
\cite{pap1}. In that paper we showed that the reflected component in the
 spectrum of Cyg X-1 is less variable than the underlying continuum at the
 frequencies above $\sim$1 Hz. The similar behavior, but with less statistical
significance was found in the case of GX 339-4. Similar to the discussion in
 \cite{pap1}
we can assume here that the time variations of the reflected component in the spectrum of GX 339-4 could 
be smeared out by the finite light-crossing time of the reflector (see more
extended discussion in  \cite{freqres2}). Alternatively, the observed
behavior could be explained by non-uniformity within the 
comptonizing region. For example, if the short time scale variations 
appear in geometrically inner part of the accretion flow (hot spheroid)
 and give a rise 
to significantly weaker, if any, reflected emission than the longer 
time scale events (originating in the outer regions) then we will see no 
reflection features at high Fourier frequencies. In turn, the smaller 
amplitude of 
the reflection in the inner regions of the accretion flow might be caused by
 the  screening of the reflector from the innermost regions by the 
outer parts of spheroid.

\begin{acknowledgements}
This research has made use of data obtained through the High Energy
Astrophysics Science Archive Research Center Online Service, provided 
by the NASA/Goddard Space Flight Center. MR acknowledges partial
support from RBRF grant 97-02-16264.
\end{acknowledgements}

\end{document}